\documentstyle[11pt,aaspp4]{article}
\eqsecnum

\begin{document}

\font\twelvei = cmmi10 scaled\magstep1
       \font\teni = cmmi10 \font\seveni = cmmi7
\font\mbf = cmmib10 scaled\magstep1
       \font\mbfs = cmmib10 \font\mbfss = cmmib10 scaled 833
\font\msybf = cmbsy10 scaled\magstep1
       \font\msybfs = cmbsy10 \font\msybfss = cmbsy10 scaled 833
\textfont1 = \twelvei
       \scriptfont1 = \twelvei \scriptscriptfont1 = \teni
       \def\mit{\fam1 }
\textfont9 = \mbf
       \scriptfont9 = \mbfs \scriptscriptfont9 = \mbfss
       \def\bmit{\fam9 }
\textfont10 = \msybf
       \scriptfont10 = \msybfs \scriptscriptfont10 = \msybfss
       \def\bmsy{\fam10 }

\def\ea{{\it et al.}~}
\def\etal{{\it et al.~}}
\def\eg{{\it e.g.,~}}
\def\ie{{\it i.e.,~}}

\def \msol {\rm{M}$_\odot$}
\def \mdot {\rm{M}$_\odot$~yr$^{-1}$}
\def \kms{km~$\rm{s}^{-1}$}
\def \cc{$\rm{cm}^{-3}$}
\def \arcs{\char'175}
\def \lam{$\lambda$}
\def \micra{$\mu$m}

\def\lsim{\raise0.3ex\hbox{$<$}\kern-0.75em{\lower0.65ex\hbox{$\sim$}}}
\def\gsim{\raise0.3ex\hbox{$>$}\kern-0.75em{\lower0.65ex\hbox{$\sim$}}}

\title{Laboratory Astrophysics and Collimated Stellar Outflows: The Production of
Radiatively Cooled Hypersonic Plasma Jets. \altaffilmark{5}}

\author{ S.V. Lebedev\altaffilmark{1}, J.P. Chittenden\altaffilmark{1}, F.N.
Beg\altaffilmark{1}, S.N. Bland\altaffilmark{1}, A. Ciardi\altaffilmark{1}, D.
Ampleford\altaffilmark{1}, S. Hughes\altaffilmark{1}, M.G. Haines
\altaffilmark{1}}
\author{A. Frank\altaffilmark{2,3}, E.G.  Blackman\altaffilmark{2,3},T.
Gardiner\altaffilmark{2,3}}

\altaffiltext{1}{ The Blackett  Laboratory, Imperial College,
London SW7 2BW, UK \\}
\altaffiltext{2}{Department of Physics and Astronomy, University of
    Rochester, Rochester NY 14627-0171\\}
\altaffiltext{3}{Laboratory for Laser Energetics, University of
    Rochester, Rochester NY 14627-0171}

\altaffiltext{5}{Submitted to the Astrophysical Journal Letters}

\begin{abstract}

We present first results of astrophysically relevant experiments
where highly supersonic plasma jets are generated via conically
convergent flows. The convergent flows are created by
electrodynamic acceleration of plasma in a conical array of fine
metallic wires (a modification of the wire array Z-pinch).
Stagnation of plasma flow on the axis of symmetry forms a standing
conical shock effectively collimating the flow in the axial
direction. This scenario is essentially similar to that discussed
by Canto\' ~and collaborators as a purely hydrodynamic mechanism
for jet formation in astrophysical systems. Experiments using
different materials (Al, Fe and W) show that a highly supersonic
($M\sim 20$), well-collimated jet is generated when the radiative
cooling rate of the plasma is significant. We discuss scaling
issues for the experiments and their potential use for numerical
code verification.  The experiments also may allow direct
exploration of astrophysically relevant issues such as
collimation, stability and jet-cloud interactions.

\end{abstract}

\clearpage

\section{Introduction}

Highly collimated supersonic plasma jets are ubiquitous, 
occurring in astrophysical environments as diverse as Active Galactic
Nuclei (AGN, \cite{Leahy91}), Young Stellar Objects (YSOs,
\cite{Reipurth97}) and Planetary Nebulae (PNe,
\cite{SokLiv94,Sahai00}). 
%Astrophysical jets represent extreme
%examples of plasma phenomena and 
In many situations these jets represent high mach number flows with
remarkable collimation characteristics.  In the case of
Herbig-Haro jets, observed to emanate from newly formed stars,
typical velocities ($V \sim 200 ~km ~s^{-1}$) and temperatures ($T
\sim 5000 ~K$) yield mach numbers in excess of $M = 20$. The
length to width ratios of the HH jets can be a factor of 100 or
more (\cite{Reipurth97}). In the late stages of stellar evolution
(PNe), highly collimated flows are observed which appear to change
direction with increasing distance from the source
(\cite{Bork97}). In all cases, the ability of astrophysical jets
to remain narrow as they propagate over large scales (while
changing direction) represents just one of many critical questions
facing the field. The role of instabilities, the effect of
jet/ambient-cloud collisions, the collimating effects of magnetic
fields and the nature of high density knots within the beams are
other examples of unresolved issues in jet studies.

Jets are important not only due to their 
ubiquity but also represent excellent laboratories for the
study of important plasma astrophysics 
processes (shocks, instabilities, etc.). In
terms of clarifying issues related to the evolution of
astronomical objects, the long dynamical or ``look-back'' times,
$t_{dyn} = L_j/V_j$, in jets allow them to be used as proxies for
understanding the obscured and often unobservable central sources.

Progress in the study of astrophysical jets is hampered by the
complexity of the governing (magneto)-gasdynamical equations
(\cite{Choudhuri98}). In addition, plasma effects, such as
ambipolar diffusion (\cite{Frankea99}) or particle acceleration
(e.g. \cite{B96})
, may become important in some instances.
Because of the inherent multi-dimensional, time-dependent nature
of the problem, astrophysical jet studies have relied heavily on
the use of sophisticated numerical simulations
(\cite{Blondinea1990}, \cite{StoNor94}, \cite{Cerqueira2001},
\cite{Gardiner2001}). These simulations often include a variety of
physical processes such as magnetic fields, ionization dynamics,
chemistry and radiation transfer. There exist few, if any,
analytical solutions to the governing equations that can be used
as testbeds when these processes are coupled together in a
multi-dimensional code. In addition, the need to simultaneously
simulate both large and small scales and the need for fully 3-D
evolution means many simulation studies remain under-resolved.
Thus, in spite of significant progress made by simulation studies
finding an alternative means for studying high mach number jets,
such a laboratory experiments, would be highly beneficial to the
field.

Traditional laboratory studies of jets have, unfortunately, been
limited to lower mach number flows and have not been able to
achieve the conditions required to explore key processes such as
radiative cooling.  The advent of high energy devices used for Inertial
Confinement Fusion (High Power Lasers, Fast Z Pinches) allows
for the possibility of direct experimental studies of hypersonic
flow problems including the affect of microphysical plasma
processes. Recent laboratory studies of blast waves, shock-cloud
interactions and planetary interiors have shown the promise of
these kinds of studies (\cite{Remingtonea1999}). High mach number
jets have also been the subject of these studies (\cite{logoryea2000,Raga01}.
Recent work by
(\cite{Farleyea1999,Shigemoriea2000,Stoneea2000})using high power
lasers have produced radiative flows whose scalings appear to
allow contact with astrophysical parameter regimes including
radiative jets.

In this paper we present the first results of a series of
experiments designed to study high mach number radiative plasma
jets using fast z-pinch devices.  The purpose of this presentation is to
introduce and
validate the experimental methodology and to provide justification
for their connection to astrophysically relevant issues.  In
section 2 we describe the basic experimental set-up providing
references to the extensive literature on Z-pinch technologies
which form the basis of our studies.  Section 3 provides scaling
arguments which show how these experiments can be used in
addressing astrophysical questions.  In section 4 we provide an
overview of the results. Finally, in section 5 we present our
conclusions and discuss the potential for these kinds of studies
to directly address questions of astrophysical jet formation and
propagation.

\section{Experimental Configuration}

The schematic of our experiments is shown in Fig 1. Our studies
were conducted using a "pulsed power" or "Z-pinch" machine. A
review of these types of instruments and their diagnostic methods
can be found in \cite{Ryutovea00} and references therein.  Our
experiments utilize plasma flowing off a {\it conical} array of
wires to produce a directed flow along the array axis. Below we
briefly introduce and describe the details of the experiment,
diagnostics and the component physics.

{\bf \it Acceleration from Wire Array:} In our experiments a fast
rising current, reaching 1MA in $240 ~ns$, is applied to a conical
array of fine metallic wires. The resistive heating of the wires
by the current rapidly, ($\delta t \sim$ few nanoseconds),
converts the wires into a heterogeneous structure with dense,
practically neutral cold cores surrounded by a low density hot
coronal plasma (see e.g. \cite{KalantarHammer}). The global
magnetic field generated by the current flowing through the array
produces the net $\bf{J}\times\bf{B}$ pinching force which
accelerates coronal plasma towards the array axis. The
characteristic density of the coronal plasma near the wires is $n
\sim 10^{17} cm^{-3}$ and the measured inward streaming velocity
is $v \sim 150~km^{-1}$ (measured via laser probing diagnostics,
\cite{Lebedev99}).  The wire cores act as a reservoir of material,
allowing the process of formation and sweeping of coronal plasma
from the cores to continue for the entire duration of the current
pulse in the experiment, thus producing a quasi-stationary
converging flow of plasma. The short rise-time of the current in
this system means that current flow is along the path with the
smallest inductance (at the largest possible radius), remaining
predominantly in the vicinity of the wire cores. As a result, the
streams of the coronal plasma are virtually current-free. Thus in
the current experiments it appears (based on the value of the
current density and MHD simulations) that no dynamically
significant magnetic fields exist in the coronal plasma flows or
the jets they create.

{\bf \it Conditions on Axis:} For purely cylindrical wire array
z-pinches, stagnation of the coronal plasma flow on the array axis
forms a dense, narrow and stable {\it precursor} plasma column
with characteristic density of $\rho \sim 10^{-3} ~ - ~10^{-2} ~g
~cm^{-3}$ and electron temperature of $T \sim 50~eV$.  The column
is confined by the kinetic pressure of the plasma flow ($\rho v^2
\sim \rho k T$, \cite{Lebedev2001,Lebedev99}) from the wires. By
placing the wires in a conical array, the flow
of coronal plasma retains an axial component of momentum after
convergence on the array axis. The conical standing shock that forms
on the axis will effectively redirect the flow in the axial
direction and form a plasma jet. This process has been well
studied in astrophysical contexts. Canto and collaborators
articulated the dynamics of converging conical flows in a series
of analytical and numerical studies (\cite{Canto1988,TTCR}).
Those papers demonstrated the efficacy of jet production via
oblique shocks formed on the axis of the converging flow.  In more
recent works \cite{FrBaLi96} and \cite{MelFr97} have demonstrated
how these converging conical flows can be established in a natural
way in the context of stellar wind blown bubbles (Young Stellar
Objects, Planetary Nebulae).  In section 4 we discuss the Canto
mechanism and its relation to these experiments in more detail.

{\bf \it Conical Array Experiments and Diagnostics:} Our conical
wire array experiments were performed with $1 ~cm$ long arrays
composed of 16 fine wires. The small radius of the array was
$8~mm$, and wires were inclined at an angle of $30^\circ$ to the
array axis (Fig. 1). Three different wire materials were used: Al,
stainless steel and W of $25~\mu m$, $25~\mu m$ and $18~\mu m$
diameter, respectively. Parameters of the jets, which were ejected
above the array were measured using a laser probing system
($\lambda = 532~nm$, pulse duration $0.4~ns$) with an
interferometer and two schlieren channels, and time-resolved soft
x-ray imaging. The interferometer provided two-dimensional data on
the distribution of the electron density in the jet. The schlieren
diagnostic is sensitive to the gradients of the refractive index,
and the probing laser beam in one of the schlieren channels was
delayed by 30ns, which allowed us to measure velocity of the jet
propagation in the same experiment. A Four-frame gated soft X-ray
pinhole camera ($2~ns$ gate with $9~ns$ separation) was used to
record two-dimensional time-resolved images of the jet emission.
Spatial resolution for the laser probing was $\delta x \sim
0.1~mm$ and for the soft x-ray imaging $\delta x \sim 0.3~mm$.

\section{Scaling Relations}
The purpose of our experiments is to create jets whose properties
can be scaled to those of real astrophysical systems.  Solutions
to the hydrodynamic equations which govern jets produced in these
studies can be shown to be invariant under certain scaling
relations (\cite{LandauLif87}).  In general, when dissipative
processes can be ignored it is possible to identify key
dimensionless parameters which will control the evolution of the
system.  When the scaling relations are obtained, two systems with
the same values of these dimensionless parameters will behave
similarly regardless their differences in physical size and
temporal duration. The allowed scaling transformations between
laboratory experiments and astrophysical systems has been
articulated by \cite{Ryutovea1999} and \cite{Ryutovea2001}

In our jet studies there are 6 dimensionless parameters we need to
consider.  The first three relate to global properties of the
jets (\cite{Blondinea1990}). They are: the
sonic mach number $M=v_j/c$; the jet to ambient density ratio
$\eta = \rho_j/\rho_a$; the ratio of cooling length to jet radius
$\chi = d_c/r_j$.  Typical ranges of values obtained in stellar
jets are $M > 10$,~$\eta \ge 1$ and $\chi \le 1$.  As we shall
show in the description of results, our experiments are comfortably
within these ranges.

The second set of dimensionless parameters concerns the
microphysical properties of the flow.  The first is the
localization parameter, which is a measure of 
the fluid-like nature of the plasma relative to the scale of the jet radius:  
$\delta_{||} = \lambda_{mfp,||}/r_j$ where $\lambda_{mfp,||}$
is the mean free path of the particles parallel
to the jet radius. The second and third 
parameters concern the role of the dissipative processes:
viscosity via the Reynolds number $Re = r_jv_j/\nu$; heat
conduction via the Peclet number $Pe = r_j v_j/\nu_h$ (note $Pe$
measures the importance of heat convection in comparison with
conduction).  In these expressions $\nu$ and $\nu_h$ are the
viscosity and thermal diffusivity respectively. Using typical
conditions found in the jet beams in our experiments (downstream of the conical shock
where the plasma streams converge at the axis: $v_j \sim 200
~km/s, ~r_j \sim .5 ~mm, ~n_e \sim 10^{19} ~cm^{-3}, Z \sim 10, ~T
\sim 50 ~eV$) we find $\delta_{||} \le 10^{-4}, ~Re > 10^4, ~Pe >
10$.  
(In determining $\delta_{||}$ for the jet, we used the temperature
since the supersonic bulk flow is perpendicular to
the jet radius in the outflow. If instead we consider
$\delta_{||}$ associated with the converging flows toward the axis
which initially produces the jet, we must then use the bulk speed 
since that corresponds to the relevant velocity parallel to the jet radius. 
This would then give $\delta_{||}\sim 0.1-0.3$ for this pre-jet converging
flow.)  Note also that while $Pe > 1$ as it should be,
the value obtained is low enough such that heat conduction may
play a role in smoothing small scale features.

Thus, in general, both the localization and Reynolds
number are in the correct regime for the jet flow. 
We conclude that, with caveat that heat conduction effects must be
explored in more detail, our experiments provide conditions which
allow scalings to astrophysically appropriate environments, with the
caveat that the magnetic fields are evidently not dynamically
important inside the jets.

\section{Results}

Fig 2 shows results from laser probing measurements of a jet
formed in the tungsten wire array. These images show the high
degree of flow collimation obtained in case.  The jet has a sharp,
well-defined boundary and propagates with a velocity significantly
higher than its radial expansion. The schlieren images of the jet
shown in Fig 2 were obtained in the same experiment at two
different times, and the measured displacement of the jet tip
gives velocity $V_z \sim 200$ \kms.  Comparison of the two images
indicates a far smaller velocity of the radial expansion of the
jet, $V_r < 7$ \kms. This allows an estimate the internal Mach
number of the jet as $M \sim V_z / V_r \sim 30$. Abel inversion of
fringe shifts in the interferogram shown in Fig.2 yields typical
electron densities at the jet boundary $n \sim 10^{19}$ \cc. The
electron density in the jet decreases towards the axis. This,
however, does not necessarily mean that the mass density
distribution is also hollow. It is possible that the decrease of
the electron density near the axis reflects the difference in
ionization levels.  Note that since the jets propagate into a
vacuum in these experiments we have density ratio $\eta >> 1$

Variation of the wire material in the experiments allows us to
examine the importance of radiation losses on jet collimation by
converging conical flows. Previous experiments
(\cite{Lebedev2000,Lebedev99}) with cylindrical arrays indicate
that the parameters of the plasma flow (an inward velocity and the
mass flux) are essentially insensitive to the material used. The
increase of the atomic number (A), however, leads to a significant
increase in the rate of energy loss through radiation. The rate of
radiative cooling can be estimated using a steady state coronal
equilibrium model by \cite{Postea77}, who provided cooling tables,
$C_A(T)$, for different elements. We note that radiation in lines
is dominant in the total energy losses. In our experiments we vary
the material composing the wires using either aluminium (Al),
stainless steel (Fe) or tungsten (W) wires. Consideration of the
\cite{Postea77} results show that for these materials $C_A(T)$ has
a single peak in the range $T<100 eV$ with the peak values
$7\times 10^{-19}$,~$3\times 10^{-18}$ and $>>5 \times10^{-18}
~erg ~cm^{3}/s$ respectively for Al, Fe and W. Consideration of
the cooling parameter $\chi$ shows $d_c \propto t_c \propto T/(n_i
C_A(T))$.  Thus we expect that $\chi \le 1$ for strong shocks
which form in the flow we produce in these experiments.  Note that
$\chi$ will decrease with increasing atomic number A. We expect
the jets from tungsten wires, in particular, will be in the
strongly radiative regime.

In cylindrical wire array experiments the use of different
materials, (and hence different cooling rates), resulted in a
difference in equilibrium radius of the precursor plasma column.
The equilibrium radius was significantly smaller for the high Z
elements (\cite{Lebedev2001,Chittenden01}). For conical array
experiments the difference in cooling rates will lead to
differences in jet collimation properties.  In their description
of the theory of converging conical flows, \cite{Canto1988}
derived relations between the converging flow properties and the
resulting jet characteristics.  Of particular importance for the
results presented here they found relations between the converging
flow angle of incidence ($\theta$), jet radius ($r_j$) and the
half-opening angle $\alpha$ of the conical shock on the axis. Both
$r_j$ and $\alpha$ depend on the inverse compression ratio ($\zeta
= \rho_0/\rho_1$) behind the shock (where subscripts 0 and 1 refer
to upstream and downstream conditions).

\begin{equation}
\tan\alpha =  \frac{(1-\zeta)-\sqrt{(1-\zeta)^2 - 4\zeta
\tan^2\theta}}  {2\tan\theta} \label{alpha}
\end{equation}

\begin{equation}
 r_j = y_o
\frac{\tan\alpha}{\tan\theta+\tan\alpha} \label{rj}
\end{equation}

\noindent In the latter expression $y_o$ is a measure of the width
of the converging flow.  The equations above demonstrate that as
the compression ratio increases (and $\zeta$ decreases) both the
opening angle of the axial shock and the radius of the jet will
decrease.  Since post-shock compression will depend on the degree
of radiative cooling, equations \ref{alpha} and \ref{rj} predict
that jets formed in our experiments will become more narrow as the
atomic number associated with the plasma ions increases.  Note the
opening angle of the jet depends on the post-shock temperature and
so we expect that it will also decrease with increased cooling
(and increased atomic number).

Typical images of jets formed in experiments using conical arrays
of aluminium, stainless steel and tungsten wires are shown in Fig
3. These images demonstrate that the degree of jet collimation
does strongly depend on the atomic number. Increasing collimation
is seen for the elements with higher A.  The diameters of the jets
at the end of the formation region, where the converging plasma
flow is still present, decrease with increasing atomic number,
being $~3.5$, $3$ and $2.5 ~mm$ for Al, Fe and W, respectively.
For Al the jet is slightly diverging, while for stainless steel
the observed opening angle is close to zero, and the jet radius
remains constant over the propagation length of $l \sim 15~mm$.
For tungsten, an apparent convergence of the jet occurs at the jet
head. The radius of the jet decreases towards the tip and an
opening angle becomes negative there.  The shape of the tip may
reflect the temporal history of the formation of conical standing
shock however more study of this point is required. The observed
dependence of jet collimation on atomic number is consistent with
predictions of Canto's model and indicates that changes in jet
radius/opening angle are due to a higher levels of radiative
losses in high A plasmas.

Experimental evidence of fast radiative cooling of the jet is also
seen from gated soft x-ray images (not shown), filtered to
transmit radiation in the interval $~190-290 ~eV$. The data show
that emission from the jets is rapidly decreasing along the beam.
The length of the emitting part of the jet is significantly
smaller than that measured at the same time by laser probing for
all materials tested in the experiments (Al, Fe and W). The
characteristic length of the emission decay (from the base of the
jet) is $l_e \sim 7~mm$ for Al, $l_e \sim 5~mm$ for Fe and $l_e
\sim 2~mm$ for W, while the characteristic lengths of the jets,
seen in laser probing images at the same time, are $L_j \sim
1.5~cm$. For a $50 ~eV$ plasma we estimate that $\approx 10\%$ of
the total radiation is in this passband.

Finally we note that previous astrophysical studies of converging
conical flows have been plagued by issues of stability. All
studies to date, both numerical (\cite{TTCR,MelFr97}) and
analytical (\cite{Canto1988,FrBaLi96}) have imposed cylindrical
symmetry as a means to facilitate calculations. While these
studies demonstrated that collimation would occur, it has not been
clear if the collimation point (the conical shock) on the axis
would be stable (\cite{GarciaSegura97}).  The experimental studies
reported here shed light on this issue. It is important to
recognize that the converging plasma flows formed in our
experimental system have a substantial degree of modulation in
azimuthal direction, determined by the number of wires in the
array ($16$ in the present experiments).  Thus one expects
perturbations from axisymmetric converging conical flows with
angular wavelength of $\delta\phi \sim  23^o$. In spite of these
modulations, the degree of the jet collimation seen in the
experiments was remarkably high. To further examine stability
considerations we have performed additional experiments in which
we deliberately introduced significant axial asymmetry in the flow
by removing $2$ neighboring wires from the array. For the highly
radiative case (W),a well-collimated jet was still generated
though it emerged inclined from the axis of symmetry. These
results suggests that a high degree of symmetry in conical flow
models of proto-stellar and proto-planetary jets is not required
if radiation losses are significant in the formation region.

Finally we note that flows formed by the present technique could
be used to study propagation of astrophysical jets through ambient
clouds, a situation observed in YSOs (\cite{Reipurth96}). This
opportunity is illustrated by the image shown in Fig.4. In these
proof-of-concept experiments, a thin plastic foil was installed
perpendicular to the direction of the jet propagation. Soft x-ray
emission from the conical standing shock on the array axis is
absorbed in the foil converting its surface layer into plasma,
which expands towards the jet. The gated soft x-ray image in Fig 4
shows emission from the jet, which rapidly decays due to radiative
cooling as the jet leaves the region of formation. X-ray emission
is seen again from the region where the jet is decelerated by the
foil as its bulk energy of motion is converted into thermal
energy. This data is consistent with the conclusion that the
thermal energy of the jet during propagation is significantly
smaller than the directed kinetic energy, \ie the Mach number
of the jet is high.

\section{Conclusions and Astrophysical Implications}

We have presented results from the first of a series of
experiments designed to create and study high mach number
collimated plasma flows.  Our experiments were carried out on a
pulsed power machine which drives 1MA current through 16-wire
array Z-pinch load. The wires are configured in a conical array
such that as plasma is draw off the vaporized wires (by lorentz
forces), it streams towards the axis at an oblique angle creating
a converging conical flow.  As the flow collides on the axis, a
standing shock forms which redirects the plasma into a collimated
beam (a jet.)

Analysis of direct emission, laser probing and interferometric
measurements of the jets reveal the following characteristics.

\begin{itemize}
\item{The jets are high mach number flows with $v_j \sim 200
~km/s$ and $M_s \ge 20$.}
\item{The jets are radiative in the sense that significant thermal
energy losses to the emission of photons (predominantly in
spectral lines). We estimate values of the cooling parameter to be
$\chi = d_c/r_j \le 1$}
\item{The fluid description of the plasma is valid ($\delta =
\lambda_{mfp}/r_j < 1$), as is the inviscid assumption ($Re >> 1$)
and the assumption that Heat conduction may be neglected to first
order $Pe > 1$.  We note that the values $Pe$ are such that
diffusive effects may produce smoothing of small scale features.
Based on the values of all the dimensionless numbers, a scaling
between the behavior of the jets observed in the laboratory and
astrophysical jets should be possible.}
\item{The collimation of the flows depends strongly in the
efficacy of radiative cooling.  As radiative cooling increases the
jet radius and opening angle are observed to decrease.  These
measurements are consistent with the theory of \cite{Canto1988}.}
\item{The formation of jets via converging conical flows is stable
to azimuthal perturbations.}
\item{Our experimental configuration allows direct experimental
investigation of shock/cloud interactions.}
\end{itemize}

\noindent The results of these simulations are promising in that
they may provide a means investigating fundamental issues relating
to the collimation, propagation and stability of radiative plasma
jets.  The application of scaling relations between the laboratory
experiments and astrophysical environments (particularly stellar
jets \ie YSOs and PNe) implies that these studies can be used to
evaluate questions concerning the nature and evolution of
collimated plasma flows in space.  The experiments detailed in
this study both complement and extend previous work in this area
(\cite{Farleyea1999,Shigemoriea2000,Stoneea2000}) but provide an
entirely different rout to creating radiative scalable plasma jets
(Z-pinch vs. intense lasers).  The larger size of the jets created
in the Z-pinch systems, (scales of centimeters vs hundreds of
microns), also facilitates imaging of the resultant plasma
structures.

The use of high energy density laboratory experiments to study
cosmic plasma environments is, essentially, a new undertaking
(\cite{Remingtonea1999}). As such, the burden of proof will rest
on the practitioners of the field to demonstrate the relevancy and
effectiveness of this approach to astronomy.  The critical issue
is then to identify critical domains where the laboratory plasma
studies can directly add information to our understanding of open
astrophysical questions.  The validation of astrophysical
simulation codes using the experiments is one means of
demonstrating relevance.  It is hoped however that fundamentally
new knowledge concerning astrophysical systems can be gleaned from
these kinds of investigations.

High mach number radiative jets has already shown some promise as
being one domain where laboratory studies could prove relevant. In
particular, fundamental issues relating to the nature of high mach
number radiative jet/ambient interactions have never been explored
experimentally. The mode structure of unstable jets also remains
questions which would benefit from direct investigation. The
studies described in this letter take an important first step
towards demonstrating relevancy since they already appear to
address one long standing issue in astrophysical jet studies: the
stability of jet production via converging conical flows. While
MHD effects may be the means by which winds are launched and
collimated from stellar sources
(\cite{Pud91,Shuea94,Blackmanea00}), numerical studies have
shown that converging flows are likely to facilitate the creation
of jets even when fields are present (\cite{Gardinerea2001}). 

Thus while many issues remain to be studied in greater detail 
in future experiments, such as dynamically important magnetic fields,
present results already suggest that 
astrophysically relevant investigations of radiative plasma jets 
can be carried out with Z-pinch laboratory studies.

\acknowledgments The authors gratefully acknowledge many interesting 
discussions with Bruce Remington, Paul Drake and Dimitri Ryutov. 
We also wish to thank Alexei Poludenko, Jim Knauer and
David Meyerhofer for their help in contributing to this paper.
Support for this work was provided at the University of Rochester
by NSF grant AST-9702484, NASA grant NAG5-8428, DOE
DE-FG02-00ER54600 and the Laboratory for Laser Energetics.

\epsscale{.6}  \figcaption[]{Schematic of the
experiment. }
\epsscale{.6}  \figcaption[]{ Time-resolved
laser probing images show well-collimated plasma jet ejected from
the tungsten wire array.  Interferometry (left image) give
electron density of the jet of $~10^{19} cm^{-3}$. Two schlieren
images obtained in the same experiment with time delay of $30~ns$
give measurement of the jet velocity of $200$ km/s.}
\epsscale{.6}  \figcaption[]{Laser probing
images of plasma jets formed in Al, stainless steel and W wire
arrays show that degree of collimation increases for elements with
higher atomic number, in which rate of radiative cooling is
higher. }
\epsscale{.6}  \figcaption[]{Time-resolved soft
x-ray image showing interaction of the jet with thin plastic foil.
Emission from the jet rapidly decays due to radiative cooling as
the jet leaves formation region and emission is seen again from
the region where the jet is decelerated by the foil and kinetic
energy is converted into thermal energy.}

\end{document}